\documentclass[12pt]{article}
\usepackage{amssymb,amsmath,epsfig}

\begin{document}

\title{\bf Charged Cylindrical Collapse of Anisotropic Fluid}
\author{M. Sharif \thanks {msharif@math.pu.edu.pk} and Sundas
Fatima
\thanks {sundas\_pu@yahoo.com}\\
Department of Mathematics, University of the Punjab,\\
Quaid-e-Azam Campus, Lahore-54590, Pakistan.}

\date{}

\maketitle
\begin{abstract}
Following the scheme developed by Misner and Sharp, we discuss the
dynamics of gravitational collapse. For this purpose, an interior
cylindrically symmetric spacetime is matched to an exterior charged
static cylindrically symmetric spacetime using the Darmois matching
conditions. Dynamical equations are obtained with matter dissipating
in the form of shear viscosity. The effect of charge and dissipative
quantities over the cylindrical collapse are studied. Finally, we
show that homogeneity in energy density and conformal flatness of
spacetime are necessary and sufficient for each other.
\end{abstract}

{\bf Keywords:} Gravitational collapse; Dissipation; Junction
conditions; Dynamical equations; Weyl tensor.

\section{Introduction}

A significant problem in gravitation theory and relativistic
astrophysics is to understand the final fate of an endless
gravitational collapse. A massive star undergoes a continual
gravitational collapse at the end of its life cycle. This happens
when a star has exhausted its nuclear fuel that provided a balance
against the internal pull of gravity.

The importance of gravitational collapse in relativistic
astrophysics was realized with the pioneer work of Oppenheimer and
Snyder \cite{1}. They used general relativity to study the dynamical
collapse of a homogenous spherical dust cloud under its own gravity.
Joshi and Singh \cite{2} explored the spherically symmetric collapse
of an inhomogeneous dust cloud. They found the end state of the
gravitational collapse as a black hole or a naked singularity
depending upon the initial density distribution and the radius of
massive body. One may, however, consider dust as somewhat
unrealistic form of matter, especially towards the end stages of a
collapse, when pressures should be important. Keeping this fact in
mind, the gravitational collapse of a perfect fluid and more general
forms of matter have been studied. Misner and Sharp \cite{3}
discussed the gravitational collapse by taking the spherically
symmetric ideal fluid in the interior and the Schwarzschild
spacetime in the exterior of a star. They provided a full account of
the dynamical equations governing the adiabatic relativistic
collapse.

Darmois \cite{4} presented junction conditions that joined two
solutions of the Einstein field equations across the surfaces of
discontinuity. Sharif and Ahmad \cite{5} discussed junction
conditions between static exterior and non-static interior spacetime
in the presence of a positive cosmological constant. They also
investigated the effect of a positive cosmological constant on
spherically symmetric collapse with a perfect fluid. It was
concluded that a positive cosmological constant slows down the rate
of collapse. The same authors \cite{6} also worked on cylindrical
collapse of two perfect fluids using high speed approximation scheme
and examined the effects of pressure on the high speed collapse for
two possible cases. Kurita and Nakao \cite{7} discussed the collapse
of null dust in the cylindrically symmetric spacetime and found a
naked singularity at the symmetric axis.

Gravitational collapse is a highly dissipative process
\cite{12}-\cite{12b} whose effects cannot be ignored in the study of
collapse. Chan \cite{11} studied a realistic model for a radiating
star which undergoes dissipation in the form of radial heat flow and
shear viscosity. He concluded that shear viscosity would increase
anisotropy of pressure and also plays an important role in the study
of gravitational collapse. The assumption of shear free motion of
the fluid \cite{11a}-\cite{11c} is used to obtain exact solutions of
the field equations but it is an unrealistic approach.

Herrera and Santos \cite{12a} studied dynamical description of
gravitational collapse in view of Misner and Sharp's formulation.
Matter under consideration was distributed with spherical symmetry
and energy loss in the form of heat flow and radiation. Herrera et
al. \cite{15} formulated the set of equations with regularity and
matching conditions for the static cylindrically symmetric
distribution of matter. They showed that any conformally flat
cylindrically symmetric static source cannot be matched to the
Levi-Civita spacetime by using Darmois junction conditions. One of
the authors (LH) \cite{14} discussed the inertia of heat and its
role in the dynamics of dissipative collapse. Herrera et al.
\cite{16} also formulated the dynamical equations to include
dissipation in the form of heat flow, radiation, shear and bulk
viscosity and then coupled with causal transport equations.
Recently, Sharif and Rehmat \cite{17} extended this work for the
plane symmetric gravitational collapse.

Some literature indicates keen interest for the inclusion of an
electromagnetic field to discuss gravitational collapse. Bekenstein
\cite{8} generalized the Oppenheimer-Volkoff equations of
hydrostatic equilibrium \cite{8a} and Misner-Sharp formulation for
the dynamics of spherical gravitational collapse to the charged
case. Nath et al. \cite{9} explored gravitational collapse in the
presence of electromagnetic field by using the junction conditions
between quasi-spherical Szekeres spacetime in the interior and the
charged Vaidya spacetime in the exterior region. They concluded that
formation of a naked singularity was enhanced by an electromagnetic
field. Sharif and Abbas \cite{10} investigated the effect of an
electromagnetic field on the spherically symmetric collapse with the
perfect fluid in the presence of positive cosmological constant.

In a recent paper, Di Prisco et al. \cite{13} derived dynamical
equations for the spherically symmetric collapse by including an
electromagnetic field. They concluded that Coulomb repulsion might
prevent the gravitational collapse of the sphere. They also found
the effect of charge on the relation between the Weyl tensor and the
inhomogeneity of energy density. In this paper, we study the
dynamics of a charged cylindrically symmetric spacetime to see the
effect of charge on the rate of gravitational collapse.

The format of the paper is the following. In the next section, we
describe the gravitational source and some physical quantities. The
Einstein-Maxwell field equation are given in section \textbf{3} and
junction conditions are derived in section \textbf{4}. We formulate
the dynamical equations in section \textbf{5} and the relation
between the Weyl tensor and the density homogeneity is given in
section \textbf{6}. The last section sums up the main results of the
paper.

\section{Interior Matter Distribution and Some Physical Quantities}

We consider a cylindrical surface with its motion described by a
timelike three surface $\Sigma$, which divides $4D$ spacetime into
interior $M^{-}$ and exterior $M^{+}$ manifolds. We assume co-moving
coordinates inside hypersurface $\Sigma$. The interior cylindrically
symmetric metric is given by
\begin{eqnarray}\label{1}
&&ds^{2}_{-}=-A^{2}dt^{2}+B^{2}dr^{2}+C^{2}(d{\phi}^{2}+dz^{2}),\nonumber\\
&&-\infty<t<+\infty,\quad 0\leqslant {r} <+\infty,\quad
0\leqslant\phi\leqslant{2}\pi,\quad -\infty<z<+\infty
\end{eqnarray}
where ${\{\chi^{-\mu}\}}\equiv\{t,r,\phi,z\}~(\mu=0,1,2,3)$ and $A$,
$B$ and $C$ are functions of $t$ and $r$. Matter under consideration
is anisotropic fluid which undergoes dissipation in the form of
shear viscosity. The energy-momentum tensor for such a fluid is
defined as
\begin{equation}\label{2}
T_{\alpha\beta}=(\mu+P_{\bot})V_{\alpha}V_{\beta}+P_{\bot}g_{\alpha\beta}+(P_{r}-P_{\bot})\chi_{\alpha}
\chi_{\beta}-2\eta\sigma_{\alpha\beta},
\end{equation}
where $\mu,~P_{r},~P_{\bot},~\eta,~V_{\alpha}$ and $\chi_{\alpha}$
are the energy density, the radial pressure, the tangential
pressure, the coefficient of shear viscosity, the four-velocity of
the fluid and the unit four-vector along the radial direction
respectively. These quantities satisfy
\begin{eqnarray}\label{3}
V^{\alpha}V_{\alpha}=-1,\quad \chi^{\alpha}\chi_{\alpha}=1,\quad
\chi^{\alpha}V_{\alpha}=0.
\end{eqnarray}

The shear tensor $\sigma_{ab}$ is defined by
\begin{equation}\label{4}
\sigma_{\alpha\beta}=V_{(\alpha;\beta)}+a_{(\alpha}V_{\beta)}-\frac{1}{3}\Theta
(g_{\alpha\beta}+V_{\alpha}V_{\beta}),
\end{equation}
where the acceleration $a_{a}$ and the expansion $\Theta$ are
given by
\begin{equation}\label{5}
a_{\alpha}=V_{\alpha;\beta}V^{\beta},\quad
\Theta=V^{\alpha}_{;\alpha}.
\end{equation}
Let it be mentioned here that the bulk viscosity does not appear
explicitly as it has been absorbed in the form of radial and
tangential pressures of the collapsing fluid. The four-velocity and
the unit four-vector are given by
\begin{equation}\label{6}
V^{\alpha}=A^{-1}\delta^{\alpha}_{0},\quad
\chi^{\alpha}=B^{-1}\delta^{\alpha}_{1}.
\end{equation}
From Eqs.(\ref{4}) and (\ref{6}), the non-zero components of the
shear tensor are
\begin{equation}\label{7}
\sigma_{11}=\frac{2}{\sqrt3}{B^{2}}\sigma,\quad
\sigma_{22}=\sigma_{33}=-\frac{1}{\sqrt3}{C^{2}}\sigma.
\end{equation}
The shear scalar $\sigma$ is defined by \cite{18}
\begin{equation}\label{8}
\sigma=\frac{1}{\sqrt{3}A}\left(\frac{\dot{B}}{B}-\frac{\dot{C}}{C}\right),
\end{equation}
where dot refers to differentiation with respect to $t$. Thus we
have
\begin{equation}\label{9}
\sigma_{\alpha\beta}\sigma^{\alpha\beta}=2 \sigma^{2}.
\end{equation}
Using Eqs.(\ref{5}) and (\ref{6}), it follows that
\begin{equation}\label{10}
a_{1}=\frac{A'}{A},\quad
\Theta=\frac{1}{A}\left(\frac{\dot{B}}{B}+2\frac{\dot{C}}{C
}\right),
\end{equation}
where prime represents derivative with respect to $r$.

The C-energy for the cylindrical symmetric spacetime is defined by
\cite{19}
\begin{eqnarray}\label{11}
E=\frac{1}{8}(1-l^{-2}\nabla^{a}r\nabla_{a}r),
\end{eqnarray}
where
\begin{eqnarray*}
\quad \rho^{2}=\xi_{(1)a}\xi^{a}_{(1)},\quad
l^{2}=\xi_{(2)a}\xi^{a}_{(2)},\quad r={\rho}l.
\end{eqnarray*}
Here $\rho$ is the circumference radius, $l$ is the specific length,
$r$ is the areal radius, $\xi^{a}$ stands for Killing vectors of
cylindrically symmetric spacetime and $E$ represents the
gravitational energy per specific length of the cylinder. Thus the
specific energy of the cylinder with the contribution of
electromagnetic field in the interior region can be written as
\cite{20}
\begin{eqnarray}\label{12}
E'=\frac{l}{8}+\frac{C}{2}\left(\frac{\dot{C}^{2}}{A^{2}}
-\frac{C'^{2}}{B^{2}}\right)+\frac{s^{2}}{2C}.
\end{eqnarray}

\section{The Field Equations}

The Maxwell equations are given by
\begin{eqnarray}\label{14}
F_{\alpha\beta}&=&\phi_{\beta,\alpha}-\phi_{\alpha,\beta},\\\label{15}
{F^{\alpha\beta}}_{;\beta}&=&\mu_{0}J^{\alpha},
\end{eqnarray}
where $F_{\alpha\beta}$ is the Maxwell field tensor, $\phi_{\alpha}$
is the four-potential and $J_{\alpha}$ is the four-current. We can
write the electromagnetic energy-momentum tensor in the form
\begin{equation}\label{13}
E_{\alpha\beta}=\frac{1}{4\pi}\left(F_{\alpha}^{\gamma}F_{\beta\gamma}
-\frac{1}{4}F^{\gamma\delta}F_{\gamma\delta}g_{\alpha\beta}\right).
\end{equation}
It is assumed that the charge is at rest with respect to the
co-moving coordinates, thus the magnetic field is zero.
Consequently, the four potential and the four current will become
\begin{equation}\label{16}
\phi_{\alpha}=\Phi{\delta^{0}_{\alpha}},\quad
J^{\alpha}={\rho}V^{\alpha},
\end{equation}
where $\Phi=\Phi(t,r)$ is an arbitrary function and $\rho=\rho(t,r)$
is the charge density. The charge conservation yields
\begin{equation}\label{17}
s\left(r\right)=2{\pi}{\int^{r}_{0}}{\rho}BC^{2}dr,
\end{equation}
where $s(r)$ is the total electric charge of the interior. For the
interior spacetime, using Eqs.(\ref{6}) and (\ref{16}), the Maxwell
equations take the following form
\begin{eqnarray}\label{18}
\Phi''-\left(\frac{A'}{A}+\frac{B'}{B}-2\frac{C'}{C}\right){\Phi'}&=&{\mu_{0}}{\rho}AB^{2},
\\\label{19}
{\dot{\Phi}}'-\left(\frac{\dot{A}}{A}+\frac{\dot{B}}{B}-2\frac{\dot{C}}{C}\right){\Phi'}&=&0.
\end{eqnarray}
Solving these equations simultaneously, it follows that
\begin{equation}\label{20}
\Phi'=\frac{\mu_{0}sAB}{2{\pi}C^{2}}.
\end{equation}

The Einstein field equations for the interior metric can be written
\begin{eqnarray}\label{21}
G_{\alpha\beta}=8\pi(T_{\alpha\beta}+E_{\alpha\beta}).
\end{eqnarray}
Using Eqs.(\ref{1}), (\ref{2}), (\ref{6}), (\ref{13}) and
(\ref{20}), we can write the nonvanishing components as follows
\begin{eqnarray}\label{22}
8{\pi}(T_{00}+E_{00})&=&8{\pi}{\mu}A^{2}+\frac{{s^{2}}{\mu^{2}_{0}}
{A^{2}}}{4{\pi^2}{C^{4}}}\nonumber\\
&=&\frac{\dot{C}}{C}\left(2\frac{\dot{B}}{B}+\frac{\dot{C}}{C}\right)
+\left(\frac{A}{B}\right)^{2}\left(-2\frac{C''}{C}
+\frac{C'}{C}\left(2\frac{B'}{B}-\frac{C'}{C}\right)\right),\nonumber\\
\\\label{23}
8{\pi}(T_{01}+E_{01})&=&0=-2\left(\frac{\dot{C'}}{C}-\frac{\dot{B}C'}{BC}-\frac{\dot{C}A'}{CA}\right),
\\\label{24}
8{\pi}(T_{11}+E_{11})&=&8{\pi}\left(P_{r}-\frac{4}{\sqrt{3}}{\eta}{\sigma}\right)B^{2}
-\frac{s^{2}{\mu^{2}_{0}}B^{2}}{4{\pi^2}{C^{4}}}\nonumber\\
&=&-\left(\frac{B}{A}\right)^{2}\left(2\frac{\ddot{C}}{C}+\left(\frac{\dot{C}}{C}\right)^{2}
-2\frac{\dot{A}\dot{C}}{AC}\right)+\left(\frac{C'}{C}\right)^{2}+2\frac{A'C'}{AC},\nonumber\\
\\\label{25}
8{\pi}(T_{22}+E_{22})&=&8{\pi}\left(P_{\bot}+\frac{2}{\sqrt{3}}{\eta}{\sigma}\right)C^{2}
+\frac{s^{2}{\mu^{2}_{0}}}{4{\pi^2}C^{2}}\nonumber\\
&=&-\left(\frac{C}{A}\right)^{2}\left(\frac{\ddot{B}}
{B}+\frac{\ddot{C}}{C}-\frac{\dot{A}}{A}\left(\frac{\dot{B}}{B}+\frac{\dot{C}}{C}\right)+
\frac{\dot{B}\dot{C}}{BC}\right)\nonumber\\
&+&\left(\frac{C}{B}\right)^{2}\left(\frac{A''}
{A}+\frac{C''}{C}-\frac{A'}{A}\left(\frac{B'}{B}-\frac{C'}{C}\right)-\frac{B'C'}
{BC}\right).
\end{eqnarray}
Equation (\ref{23}) can be re-written in the following form
\begin{eqnarray}\label{25-a}
\frac{1}{3}(\Theta-{\sqrt3}\sigma)'-{\sqrt3}\sigma\frac{C'}{C}=0.
\end{eqnarray}

\section{Junction Conditions}

In this section, we formulate the junction conditions for the
interior and exterior manifolds. The interior manifold is given by
Eq.(\ref{1}) and the exterior manifold is the charged static
cylindrically symmetric spacetime given by \cite{21}
\begin{eqnarray}\label{26}
ds^{2}_{+}=-NdT^{2}+\frac{1}{N}dR^{2}+R^{2}(d{\phi}^{2}+dz^{2}),
\end{eqnarray}
where
\begin{equation*}
N(R)=\left(\frac{Q^{2}}{R^{2}}-2\frac{M}{R}\right)
\end{equation*}
and $\chi^{+\mu}\equiv \{T,R,\phi,z\}$. We can write the metric for
the hypersurface $\Sigma$ in the following form
\begin{equation}\label{27}
(ds^{2})_{\Sigma}=-{d{\tau}}^2+A^{2}(\tau)(d\phi^{2}+dz^{2}),
\end{equation}
where $\xi^{i}\equiv(\tau,\phi,z)~(i=0,2,3)$ represent the intrinsic
coordinates of $\Sigma$.

The Darmois junction conditions \cite{4} can be stated as follows:
\begin{itemize}
\item The continuity of the first fundamental form. This implies the continuity of
the metrics over the hypersurface
\begin{equation}\label{28}
(ds^{2})_{\Sigma}=(ds^{2}_{-})_{\Sigma}=(ds^{2}_{+})_{\Sigma}.
\end{equation}
\item  The continuity of the second fundamental form. This gives the continuity
of the extrinsic curvature $K_{ij}$ over the hypersurface
\begin{equation}\label{29}
[K_{ij}]=K^{+}_{ij}-K^{-}_{ij}=0.
\end{equation}
\end{itemize}
$K^{\pm}_{ij}$ is the extrinsic curvature given by
\begin{equation}\label{30}
K^{\pm}_{ij}=-n^{\pm}_{\sigma}(\frac{{\partial}^2\chi^{\sigma}_{\pm}}
{{\partial}{\xi}^i{\partial}{\xi}^j}+{\Gamma}^{\sigma}_{{\mu}{\nu}}
\frac{{{\partial}\chi^{\mu}_{\pm}}{{\partial}\chi^{\nu}_{\pm}}}
{{\partial}{\xi}^i{\partial}{\xi}^j}),\quad({\sigma},
{\mu},{\nu}=0,1,2,3).
\end{equation}
where $n^{\pm}_{\sigma}$ are the components of outward unit normal
to the hypersurface in the coordinates $\chi^{{\pm}\mu}$.

We can write the equations of the hypersurface as follows:
\begin{eqnarray}\label{31}
f_{-}(t,r)&=&r-r_{\Sigma}=0,\\
f_{+}(T,R)&=&R-R_{\Sigma}(T)=0,\label{32}
\end{eqnarray}
where $r_{\Sigma}$ is a constant. Using Eqs.(\ref{31}) and
(\ref{32}) in Eqs.(\ref{1}) and (\ref{26}), we have the interior and
exterior spacetimes on $\Sigma$ respectively
\begin{eqnarray}\label{33}
(ds^{2}_{-})_{\Sigma}&=&-A^{2}(t,r_{\Sigma})dt^{2}+C^{2}(t,r_{\Sigma})(d\phi^{2}+dz^{2}).\\
\label{34}
(ds^{2}_{+})_{\Sigma}&=&-\left[N(R_{\Sigma})-(N(R_{\Sigma}))^{-1}
\left(\frac{dR_{\Sigma}}{dT}\right)^{2}\right]dT^{2} +R^{2}_{\Sigma}
(d\phi^{2}+dz^{2}).\nonumber\\
\end{eqnarray}
The continuity of the first fundamental form gives
\begin{eqnarray}\label{35}
R^{2}_{\Sigma}&=&C(t,r_{\Sigma}),\\ \label{36}
\frac{dt}{d\tau}&=&\frac{1}{A},\\ \label{37}
\frac{dT}{d\tau}&=&\left[N(R_{\Sigma})-{N(R_{\Sigma})}^{-1}
\left(\frac{dR_{\Sigma}}{dT}\right)^{2}\right]^{-\frac{1}{2}}.
\end{eqnarray}
Now we consider the second fundamental form over $\Sigma$. For this
purpose, we need the outward unit normals to $\Sigma$ using
Eqs.(\ref{31}) and (\ref{32}),
\begin{eqnarray}\label{38}
n^{-}_{\mu}&=&B(0,1,0,0),\\ \label{39}
n^{+}_{\mu}&=&\left[N(R)-(N(R))^{-1}\left(\frac{dR}{dT}\right)^{2}\right]^{-\frac{1}{2}}
\left(-\frac{dR}{dT},1,0,0\right).
\end{eqnarray}

The surviving components of the extrinsic curvature $K^{\pm}_{ij}$
can be given as follows
\begin{eqnarray}\label{40}
K^{-}_{00}&=&-\left(\frac{A'}{AB}\right)_{\Sigma},\\ \label{41}
K^{+}_{00}&=&\left[\frac{d^{2}T}{d\tau^{2}}\frac{dR}{d\tau}-
\frac{d^{2}R}{d\tau^{2}}\frac{dT}{d\tau}-\frac{N}{2}\frac{dN}{dR}\left(\frac{dT}
{d\tau}\right)^{3}+{\frac{3}{2N}}{\frac{dN}{dR}}\left(\frac{dR}
{d\tau}\right)^{2}{\frac{dT}{d{\tau}}}\right]_{\Sigma},\\
\label{42}
K^{-}_{22}&=&K^{-}_{33}=\left(\frac{CC'}{B}\right)_{\Sigma},\\
\label{43}
K^{+}_{22}&=&K^{+}_{33}=\left[NR\frac{dT}{d\tau}\right]_{\Sigma}.
\end{eqnarray}
The continuity of the second fundamental form, using Eq.(\ref{29}),
yields
\begin{eqnarray}\label{44}
\left[\frac{d^{2}T}{d\tau^{2}}\frac{dR}{d\tau}-
\frac{d^{2}R}{d\tau^{2}}\frac{dT}{d\tau}-\frac{N}{2}\frac{dN}{dR}\left(\frac{dT}
{d\tau}\right)^{3}+{\frac{3}{2N}\frac{dN}{dR}\left(\frac{dR}
{d\tau}\right)^{2}\frac{dT}{d\tau}}\right]_{\Sigma}\nonumber\\
=-\left(\frac{A'}{AB}\right)_{\Sigma},
\\\label{45}
\left[NR\frac{dT}{d\tau}\right]_{\Sigma}=\left(\frac{CC'}{B}\right)_{\Sigma}.
\end{eqnarray}
Making use of Eq.(\ref{37}), it follows
\begin{equation}\label{46}
\frac{dT}{d\tau}=\frac{1}{N}\sqrt{N+\left(\frac{dR}{d\tau}\right)^{2}}.
\end{equation}
Substituting  Eq.(\ref{46}) in Eq.(\ref{45}), we obtain
\begin{equation}\label{47}
M=\frac{C}{2}\left(\frac{\dot{C}^2}{A^{2}}-\frac{C'^2}{B^{2}}\right)+\frac{Q^{2}}{2C}.
\end{equation}
Thus, if the interior and the exterior charges are equal over the
hypersurface, i.e., $Q=s$, then we have
\begin{equation}\label{48}
{E'-\frac{l}{8}}\overset{\Sigma}{=}M.
\end{equation}
The difference between these two masses is equal to $\frac{l}{8}$,
which is due to the least unsatisfactory definition of Thorne
C-energy. Differentiating Eq.(\ref{46}) with respect to $\tau$, then
inserting this value in Eq.(\ref{44}) and making use of
Eq.(\ref{45}), we can write Eq.(\ref{44}) as
\begin{eqnarray}\label{49}
\frac{\dot{C}'}{C}-\frac{\dot{B}C'}{BC}-\frac{A'\dot{C}}{AC}=0
\end{eqnarray}
This equation identically satisfies Eq.(\ref{23}). For the smooth
matching of the interior and exterior metrics on hypersurface,
Eqs.(\ref{35})-(\ref{37}), (\ref{47}) and (\ref{49}) are the
necessary and sufficient conditions.

\section{Dynamical Equations}

The conservation of energy-momentum,
$(T^{\alpha\beta}+E^{\alpha\beta})_{;\beta}=0$, implies that
\begin{eqnarray}\label{51}
\left(T^{\alpha\beta}+E^{\alpha\beta}\right)_{;\beta}V_{\alpha}&=&-\frac{\dot{\mu}}{A}
-\frac{\dot{B}}{AB}\left(\mu+P_{r}-\frac{4}{\sqrt3}\eta\sigma\right)\nonumber\\
&-&\frac{2\dot{C}}{AC}\left(\mu+P_{\bot}+\frac{2}{\sqrt3}\eta\sigma\right)
=0
\end{eqnarray}
and
\begin{eqnarray}\label{52}
\left(T^{\alpha\beta}+E^{\alpha\beta}\right)_{;\beta}\chi_{a}&=&\frac{1}{B}\left(P_{r}-
\frac{4}{\sqrt3}\eta\sigma\right)'+\frac{A'}{AB}\left(\mu+P_{r}-\frac{4}{\sqrt3}\eta\sigma\right)
\nonumber\\&+&\frac{2C'}{BC}\left(P_{r}-P_{\bot}-{2}{\sqrt3}\eta\sigma\right)
-\frac{\mu_{0}^{2}ss'}{16\pi^{3}BC^{4}}=0.
\end{eqnarray}
In view of Misner and Sharp's formalism, we discuss the dynamics of
a collapsing system. We introduce the proper time derivative as
\begin{equation}\label{53}
D_{T}=\frac{1}{A}\frac{\partial}{\partial{t}}.
\end{equation}
The proper radial derivative $D_{R}$ constructed  from the
circumference radius of a cylinder inside $\Sigma$ is
\begin{equation}\label{54}
D_{R}=\frac{1}{R'}\frac{\partial}{\partial{r}},
\end{equation}
where
\begin{equation}\label{55}
R=C.
\end{equation}

The fluid velocity in the case of collapse can be defined as
\begin{equation}\label{56}
U=D_{T}(R)=D_{T}(C)<0,
\end{equation}
which must be negative in the process of collapse. Using
Eq.(\ref{56}), we can re-write Eq.(\ref{12}) as
\begin{equation}\label{57}
\tilde{E}=\frac{C'}{B}=\left[U^{2}+\frac{s^{2}}{C^{2}}-\frac{2}{C}
\left(E'-\frac{1}{8}\right)\right]^{1/2}.
\end{equation}
From Eqs.(\ref{54}) and (\ref{55}), Eq.(\ref{25-a}) can be written
as
\begin{eqnarray*}
BE\left[\frac{1}{3}D_{R}(\Theta-{\sqrt3}\sigma)-{\sqrt3}\frac{\sigma}{R}\right]=0.
\end{eqnarray*}
In non-dissipative shear free case, i.e., $\eta=\sigma=0$, this
equation takes the form
\begin{eqnarray*}
D_{R}\left(\frac{U}{R}\right)=0.
\end{eqnarray*}
This means that $U\sim R$ which describes the homologous collapse.
In view of Eqs.(\ref{12}), (\ref{23}), (\ref{24}) and (\ref{53}),
the rate of variation of the C-energy turns out to be
\begin{equation}\label{58}
D_{T}E'=-4{\pi}R^{2}\left(P_{r}-\frac{4}{\sqrt3}\eta\sigma
-\frac{1}{32{\pi}R^{2}}\right)U+\frac{s^{2}U}{R^{2}}
\left(\frac{\mu_{0}^{2}}{{8}\pi^{2}}-\frac{1}{2}\right).
\end{equation}
The first term on the right-hand side of Eq.(\ref{58}) in the case
of collapse $(U<0)$ will increase the energy of the cylinder if
\begin{eqnarray}\label{59}
P_{r}-\frac{4}{\sqrt3}\eta\sigma>\frac{1}{32{\pi}R^{2}},
\end{eqnarray}
i.e., the effective radial pressure is greater than the particular
value. This increase of C-energy is due to the work being done by
the effective radial pressure. The second term in the round brackets
describe energy leaving the system due to the Coulomb repulsive
force. Similarly, using Eqs.(\ref{12}), (\ref{22}), (\ref{23}) and
(\ref{54}), we obtain
\begin{equation}\label{60}
D_{R}E'=4{\pi}{\mu}R^{2}+\frac{1}{8}+\frac{s}{R}D_{R}s+\frac{s^{2}}{R^{2}}\left
(\frac{\mu_{0}^{2}}{{8}\pi^{2}}-\frac{1}{2}\right).
\end{equation}

This equation indicates variation of the total energy between
adjoining cylindrical surfaces inside the fluid distribution. The
first term on right hand side gives the contribution of the energy
density of the fluid element and the remaining terms are a constant
and the electromagnetic contribution respectively. Integration of
Eq.(\ref{60}) leads to
\begin{equation}\label{61}
E'={\int^{R}_{0}}4{\pi}{\mu}R^{2}dR+\frac{R}{8}+\frac{s^{2}}{2R}+
\frac{\mu_{0}^{2}}{{8}\pi^{2}}{\int^{R}_{0}}\frac{s^{2}}{R^{2}}dR.
\end{equation}
Using Eqs.(\ref{12}), (\ref{24}), (\ref{56}) and (\ref{57}), we can
obtain the acceleration $D_{T}U$ of a collapsing matter inside
$\Sigma$as follows:
\begin{equation}\label{62}
D_{T}U=-\frac{1}{R^{2}}\left(E'-\frac{l}{8}\right)
-4\pi{R}\left(P_{r}-\frac{4}{\sqrt{3}}\eta\sigma\right)+\frac{\tilde{E}A'}{AB}
+\frac{s^{2}}{R^{3}}\left(\frac{\mu_{0}^{2}}{{8}\pi^{2}}+\frac{1}{2}\right).
\end{equation}
Inserting the value of $\frac{A'}{A}$ from Eq.(\ref{62}) into
Eq.(\ref{52}), it follows that
\begin{eqnarray}\label{63}
&&\left(\mu+P_{r}-\frac{4}{\sqrt{3}}\eta\sigma\right)D_{T}U
=-\left(\mu+P_{r}-\frac{4} {\sqrt{3}}\eta\sigma\right)\nonumber\\
&&\times\left[\frac{1}{R^{2}}\left(E'-\frac{l}{8}\right)+4\pi\left(P_{r}-\frac{4}{\sqrt{3}}
\eta\sigma\right)R
-\frac{s^{2}}{R^{3}}\left(\frac{\mu_{0}^{2}}{{8}\pi^{2}}+
\frac{1}{2}\right)\right]\nonumber\\
&&-\tilde{E}^2\left[D_{R}\left(P_{r}-
\frac{4}{\sqrt{3}}\eta\sigma\right)+2\left(P_{r}-P_{\bot}
-2\sqrt{3}\eta\sigma\right)\frac{1}{R}-\frac{{\mu_{0}^{2}}s}{16{\pi^{3}}R^{4}}D_{R}s\right].\nonumber\\
\end{eqnarray}

This equation has the ``Newtonian" form, i.e.,
\begin{equation*}
Mass~density\times Acceleration=Force.
\end{equation*}
Now we analyse the terms appearing in this equation as follows: The
term in the round brackets on the left hand side represents inertial
mass density. This gives the effect of the dissipative terms but
there is no contribution of the electric charge. The remaining term
on the left hand side is acceleration. There are two main terms on
the right hand side. The first term represents the gravitational
force. The factor within the round brackets is the same as on the
left. It represents passive gravitational mass density by
equivalence principle. The factor within the first square brackets
shows how specific length, dissipation, and the electric charge
affect the active gravitational mass term. Making use of
Eq.(\ref{61}) in Eq.(\ref{63}), it turns out
\begin{equation*}\label{64}
{\int^{R}_{0}}\frac{s^{2}}{R^{2}}dR > \frac{s^{2}}{R},
\end{equation*}
which on differentiation yields
\begin{equation}\label{65}
\frac{s}{R} > D_{R}s.
\end{equation}
Thus the charge will increase the active gravitational mass only if
this condition is satisfied. This increase of active gravitational
mass causes the rapid collapse.

The second term in the second square brackets constitute
hydrodynamical forces. It consists of further three terms. The first
contribution simply represents the gradient of the total effective
radial pressure (including the influence of shear viscosity on
$P_{r}$) which is always negative and is directed outward, this
would prevent the gravitational collapse. The second term exhibits
the effect of the local anisotropy of pressure with shear viscosity.
If anisotropic pressure is positive then it increases the rate of
collapse otherwise it decreases. In the last term, we have the
Coulomb repulsion which may prevent the gravitational collapse of
the cylinder.

For hydrostatic equilibrium, i.e., $U=0,~\eta=0$, Eq.(\ref{63})
turns out to be
\begin{eqnarray}
D_{R}P_{r}&=&\frac{{\mu_{0}^{2}}s}{16{\pi^{3}}R^{4}}D_{R}s-2(P_{r}-P_{\bot})\nonumber\\
&-&\frac{C'^{2}}{B^{2}}(\mu+P_{r})
\left[\frac{1}{R^{2}}\left(E'-\frac{l}{8}\right)+4{\pi}{P_{r}}R
-\frac{s^{2}}{R^{3}}\left(\frac{\mu_{0}^{2}}{{8}\pi^{2}}+
\frac{1}{2}\right)\right].
\end{eqnarray}
We can also write Eq.(\ref{52}) for the hydrostatic equilibrium
$U=0,~\eta=0$ as follows:
\begin{eqnarray}\label{65-a}
P_{r}'+\frac{A'}{A}(\mu+P_{r}) +\frac{2C'}{C}(P_{r}-P_{\bot})
-\frac{\mu_{0}^{2}ss'}{16\pi^{3}C^{4}}=0.
\end{eqnarray}
This corresponds to the hydrostatic equilibrium for the spherically
symmetric case and also gives the generalization of the
Tolman-Oppenheimer-Volkoff equation for anisotropic charged fluid
\cite{21a}.

The static fluid leads to charged dust by taking $P_{r}=0=P_{\bot}$,
thus, it follows from Eqs.(\ref{65-a}) and (\ref{17}) that
\begin{eqnarray}\label{65-b}
\mu\frac{A'}{A}-\frac{\mu_{0}^{2}s{\rho}B}{8\pi^{2}C^{2}}=0.
\end{eqnarray}
Here $B$ and $C$ are function of $r$ and hence $B=C$ for a suitable
transformation of $r$. Eliminating $s$ from the field equations
(\ref{24}) and (\ref{25}), we get
\begin{eqnarray}\label{65-c}
B=C,\quad AB=1,\quad s^{2}=\frac{4\pi^{2}}{\mu_{0}^{2}}B'^{2}.
\end{eqnarray}
Inserting these values in in Eq.(\ref{65-b}), we obtain
\begin{eqnarray}\label{65-d}
\mu^{2}={\kappa}\rho^{2},\quad \kappa=\frac{\mu_{0}^{2}}{16\pi^{2}}.
\end{eqnarray}
We would like to mention here that such type of result was also
found by Bonnor \cite{21b} for arbitrary spacetime.

\section{The Weyl Tensor}

Here we shall explore the relation between the Weyl tensor and
density inhomogeneity. For this purpose, we define the Weyl scalar
$\mathcal{C}^{2}$ in terms of the Kretchman scalar $\mathcal{R}$,
the Ricci tensor ${R}_{\alpha\beta}$ and the curvature scalar
$\textrm{R}$, i.e.,
\begin{equation}\label{66}
\mathcal{C}^{2}=\mathcal{R}-2\textit{R}^{\alpha\beta}\textit{R}_{\alpha\beta}
+\frac{1}{3}\textrm{R}^{2}.
\end{equation}
Inserting the value of $\mathcal{R}$ from Eq.(\ref{84}) in the
appendix and making use of Eqs.(\ref{22})-(\ref{25}) in
Eq.(\ref{66}), it follows that
\begin{equation}\label{67}
\varepsilon=E'-\frac{l}{8}-\frac{4\pi}{3}R^{3}\left(\mu-P_{r}+P_{\bot}+2{\sqrt{3}}
\eta\sigma\right)-\frac{s^{2}}{R}\left(\frac{\mu_{0}^{2}}{{8}\pi^{2}}+\frac{1}{2}\right),
\end{equation}
where $\varepsilon$ is defined as
\begin{equation}\label{68}
\varepsilon=\frac{\mathcal{C}}{48^\frac{1}{2}}R^{3}.
\end{equation}
Applying the definitions of $D_T$ and $D_R$ from Eqs.(\ref{58}) and
(\ref{60}) respectively to (\ref{67}), we obtain
\begin{eqnarray}\label{69}
D_{T}{\varepsilon}&=&-4{\pi}
[\frac{1}{3}{R^{3}}D_{T}\left(\mu-P_{r}+P_{\bot}+2{\sqrt{3}}
\eta\sigma\right)\nonumber\\
&+&\left(\mu+P_{\bot}+\frac{2}{\sqrt{3}}\eta\sigma\right){UR}^{2}]
+\frac{{\mu^{2}_{0}}{s^{2}}U}{4{\pi^{2}}R^{2}}.
\end{eqnarray}
and
\begin{eqnarray}\label{70}
D_{R}{\varepsilon}&=&4{\pi}[-\frac{1}{3}{R^{3}}D_{R}\left(\mu-P_{r}+P_{\bot}+2{\sqrt{3}}
\eta\sigma\right)\nonumber\\
&+&\left(P_{r}-P_{\bot}-2\sqrt{3}\eta\sigma\right)R^{2}]+
\frac{\mu^{2}_{0}}{4{\pi^{2}}}\left[-\frac{sD_{R}s}{R}+{\left(\frac{s}{R}\right)}^{2}\right].
\end{eqnarray}
This shows that production of density inhomogeneity is directly
linked to dissipative variables and charge. For the case of zero
charge and dissipation, we have
\begin{eqnarray}\label{71}
D_{R}{\varepsilon}+\frac{4{\pi}}{3}R^{3}D_{R}{\mu}=0.
\end{eqnarray}
This implies that if $D_{R}\mu=0$ then $\mathcal{C}=0$ (using the
regular axis condition). Conversely, the conformally flat condition
implies homogeneity in the energy density. This result has already
been verified for spherically symmetric gravitational collapse
\cite{13}.

\section{Summary and Conclusion}

To investigate how the system gradually changes with time, we have
formulated a dynamical description of the cylindrically symmetric
spacetime using Misner and Sharp's approach. Dissipative effects and
anisotropic pressure have been taken into account. The junction
conditions between cylindrically symmetric in the interior and
charged static cylindrically symmetric spacetime in the exterior
provides the gravitational mass which causes gravity in the exterior
region.

We have found the behavior of charge and pressure through the
dynamical equations. It turns out that electric charge (unlike
pressure) does not always produce a regeneration effect, i.e., the
pressure trying to keep the star in equilibrium through the pressure
gradients, at the same time contributes to the active gravitational
mass. Thus it increases the gravitational attraction and hence it
promotes stellar collapse at the same time. This is due to the
inequality (63) which implies that if
\begin{equation*}
\frac{s}{R} > D_{R}s,
\end{equation*}
then it decreases the gravitational mass and also due to the Coulomb
force that always opposes the gravitational force. We would like to
mention here that our results indicate similarity with those found
for the spherically symmetric spacetime \cite{13}.

We have also established an expression indicating the relevance of
the electric charge with the Weyl tensor and density inhomogeneity.
Using the regular axis condition, it has been shown that homogeneity
in energy density and conformal flatness of spacetime are necessary
and sufficient for each other. We would like to mention here that
the Weyl tensor contains tidal forces that make the fluid more
inhomogeneous in the process of evolution. It would be interesting
to include also a heat flux and examine the corresponding transport
equations. Also, one would be interested to extend these results for
charged plane symmetric spacetime \cite{22}.

\section*{Appendix}

The interior metric has the following nonvanishing components of the
Riemann tensor
\begin{eqnarray}\label{72}
\textit{R}_{0101}&=&AA''-B\ddot{B}-\frac{A}{B}A'B'+\frac{B}{A}\dot{A}\dot{B},\\
\label{73}
\textit{R}_{0202}&=&\frac{C}{AB^{2}}\left(-\ddot{C}AB^{2}+\dot{A}\dot{C}B^{2}
+A'C'A^{2}\right),\\
\label{74}
\textit{R}_{0212}&=&\frac{C}{AB}\left(-\dot{C}'AB+\dot{C}A'B+\dot{B}C'A\right),\\
\label{75}
\textit{R}_{1212}&=&\frac{C}{A^{2}B}\left(-C''A^{2}B+\dot{C}\dot{B}B^{2}+B'C'A^{2}\right),\\
\label{76}
\textit{R}_{2323}&=&\frac{C^{2}}{(AB)^{2}}(\dot{C}^{2}B^{2}-C'^{2}A^{2}),
\end{eqnarray}
and
\begin{eqnarray}\label{77}
\textit{R}_{0202}=\textit{R}_{0303},\quad
\textit{R}_{0212}=\textit{R}_{0313},\quad
\textit{R}_{1212}=\textit{R}_{1313}.
\end{eqnarray}
Thus it has five independent components. The Kretchman scalar
$\mathcal{R}=R^{\alpha\beta\gamma\delta}R_{\alpha\beta\gamma\delta}$
becomes
\begin{eqnarray}\label{78}
\mathcal{R}=4[\frac{1}{(AB)^{4}}(\textit{R}_{0101})^{2}&+&\frac{2}{(AC)^{4}}(\textit{R}_{0202})^{2}
-\frac{4}{{(AB)^{2}}C^{4}}(\textit{R}_{0212})^{2}\nonumber\\
&+&\frac{2}{(BC)^{4}}(\textit{R}_{1212})^{2}+\frac{1}{C^{8}}(\textit{R}_{2323})^{2}].
\end{eqnarray}
The components of the Riemann tensor in terms of the Einstein tensor
and the C-energy function can be written as
\begin{eqnarray}\label{79}
\textit{R}_{0101}&=&(AB)^{2}\left[\frac{1}{2A^{2}}G_{00}-\frac{1}{2B^{2}}G_{11}+
\frac{G_{22}}{C^{2}}-\frac{2}{C^{3}}\left(E'-\frac{l}{8}-\frac{s^{2}}{2C}\right)\right],
\\\label{80}
\textit{R}_{0202}&=&(AC)^{2}\left[\frac{G_{11}}{2B^{2}}+\frac{1}{C^{3}}\left(E'-
\frac{l}{8}-\frac{s^{2}}{2C}\right)\right], \\\label{81}
\textit{R}_{0212}&=&\frac{C^{2}}{2}G_{01}, \\\label{82}
\textit{R}_{1212}&=&(BC)^{2}\left[\frac{G_{00}}{2A^{2}}-\frac{1}{C^{3}}\left(E'-
\frac{l}{8}-\frac{s^{2}}{2C}\right)\right], \\\label{83}
\textit{R}_{2323}&=&2C\left(E'-\frac{l}{8}-\frac{s^{2}}{2C}\right).
\end{eqnarray}
Inserting Eqs.(\ref{79})-(\ref{83}) into Eq.(\ref{78}), we get
\begin{eqnarray}\label{84}
\mathcal{R}&=&\frac{48}{C^{6}}\left(E'-\frac{l}{8}-\frac{s^{2}}{2C}\right)^{2}-\frac{16}{C^{3}}
\left(E'-\frac{l}{8}-\frac{s^{2}}{2C}\right)\left[\frac{G_{00}}{A^{2}}-
\frac{G_{11}}{B^{2}}+\frac{G_{22}}{C^{2}}\right]\nonumber\\
&-&{4\left(\frac{G_{01}}{AB}\right)^{2}}
+3\left[\left(\frac{G_{00}}{A^{2}}\right)^{2}+\left(\frac{G_{11}}{B^{2}}\right)^{2}\right]
+4\left(\frac{G_{22}}{C^{2}}\right)^{2}\nonumber\\
&-& 2\frac{G_{00}G_{11}}{(AB)^{2}}+
4\left(\frac{G_{00}}{A^{2}}-\frac{G_{11}}{B^{2}}\right)\frac{G_{22}}{C^{2}}.
\end{eqnarray}

\end{document}